\documentclass[12pt]{article}
\usepackage{graphicx,cite}
\usepackage{amsfonts}
\title{Lie Symmetries of Multidimensional \\
 Difference Equations}
\author{D. Levi\thanks{Dipartimento di Fisica, Universit\`a Roma Tre and
INFN--Sezione
di Roma Tre, Via della Vasca Navale 84, 00146 Rome, Italy (email:
levi@fis.uniroma3.it)}
\and S. Tremblay\thanks{Centre de Recherches Math\'ematiques and D\'epartement
 de Physique, Universit\'e de Montr\'eal, C.P. 6128, succ. Centre-ville,
Montr\'eal (QC), H3C 3J7, Canada (email: tremblay@crm.umontreal.ca)} \and P.
Winternitz\thanks{Centre de
Recherches Math\'ematiques and D\'epartement de Math\'ematiques et
Statistique, Universit\'e de Montr\'eal, C.P. 6128, succ. Centre-ville,
Montr\'eal (QC), H3C 3J7, Canada (email: wintern@crm.umontreal.ca)}}
\date{}

\def\pa {\partial}
\def\ti {\tilde}

\def\eq {\equiv}

\def\al{\alpha}
\def\b{\beta}
\def\ga{\gamma}
\def\de{\delta}

\def\ze{\zeta}

\def\la{\lambda}

\def\si{\sigma}

\def\be   {\begin{equation}}   \def\ee   {\end{equation}}
\def\ba   {\begin{array}}      \def\ea   {\end{array}}
\def\bea  {\begin{eqnarray}}   \def\eea  {\end{eqnarray}}
\def\bean {\begin{eqnarray*}}  \def\eean {\end{eqnarray*}}

\newcommand{\me}{\mathrm{e}}

\newcommand{\mpr}{\mathrm{pr}}

\newcommand{\xmn}{x_{m,n}}
\newcommand{\xmu}{x_{m+1,n}}
\newcommand{\xmd}{x_{m-1,n}}
\newcommand{\xnu}{x_{m,n+1}}
\newcommand{\xnd}{x_{m,n-1}}

\newcommand{\ymn}{y_{m,n}}
\newcommand{\ymu}{y_{m+1,n}}

\newcommand{\ynu}{y_{m,n+1}}
\newcommand{\ynd}{y_{m,n-1}}

\newcommand{\tmn}{t_{m,n}}
\newcommand{\tmu}{t_{m+1,n}}
\newcommand{\tmd}{t_{m-1,n}}
\newcommand{\tnu}{t_{m,n+1}}
\newcommand{\tnd}{t_{m,n-1}}

\newcommand{\umn}{u_{m,n}}
\newcommand{\umu}{u_{m+1,n}}
\newcommand{\umd}{u_{m-1,n}}
\newcommand{\unu}{u_{m,n+1}}

\begin{document}
\maketitle

\begin{abstract}
A method is presented for calculating the Lie point symmetries of a scalar
difference equation on a
two-dimensional lattice. The symmetry transformations act on the equations
and on the lattice. They take solutions into solutions and can be used to
perform symmetry reduction. The method generalizes one presented in a
recent publication for the case of ordinary difference equations. In turn,
it can easily be generalized to difference systems involving an arbitrary
number of dependent and independent variables.
\end{abstract}



\section{Introduction}

A recent article \cite{1} was devoted to Lie point symmetries, acting on
ordinary difference equations and lattices, while leaving their set of
solutions
invariant. The purpose of this article is to extend the previously obtained
methods and results to the case of partial difference equations, i.e. equations
involving more than one independent variable.

Algebraic techniques, making use of Lie groups and Lie algebras, have proved
themselves  to be extremely useful in the theory of differential equations
\cite{2}.

When applying similar algebraic methods to difference equations, several
decisions have to be made.

The first decision is a conceptual one. One can consider difference equations
and lattices as given objects to be studied. The aim then is to provide tools
for solving these equations, simplifying the equations, classifying equations
and their solutions, and identifying integrable, or linearizable difference
equations
\cite{1,3,4,5,10,6,7,8,lr,9,glw,19,lwm,hlw,ltw,hlw1,hl,11,12,q1,q2,q3,g,jv,h}. Alternatively, one can consider difference equations
and the lattices on which they are defined, to be auxiliary objects. They are
introduced in order to study solutions of differential equations, numerically or
otherwise. The question to be asked in this is: how does one discretize a
differential equation, while preserving its symmetry properties \cite{13,14,15,d,16}.

In this article we take the first point of view: the equation and the lattice
are {\em a priori} given. The next decision to be made is a technical one: which
aspect of symmetry to pursue. For differential equations one can look for point
symmetries, or generalized ones. When restricting to point symmetries, and
constructing the Lie algebra of the symmetry group, one can use vector fields
acting on dependent and independent variables. Alternatively and equivalently,
one can use evolutionary vector fields, acting only on dependent variables.
For difference equations, these two approaches are in general not equivalent and
may lead to different results, both of them correct and useful.

  Several aspects of symmetry for discrete equations were pursued in
earlier articles by two of the present authors (D.L. and P.W.) and collaborators. The "intrinsic method"  which provides, in an
algorithmic way, all purely point symmetries of a given differential -
difference equation on a given uniform fixed lattice was introduced in
\cite{4}. This was complemented by the "differential equations method" in
\cite{5}. In addition to point symmetries the differential equation method provides a class of generalized
symmetries. It was
pointed out that in many cases the two methods provide the same result,
i.e. all symmetries are point ones.  The two methods were successfully
applied to many specific problems \cite{5,6, glw, 19, ltw}. The advantage of these
two approaches are their simplicity, their algorithmic character, and
their close analogy to symmetries of differential equations. Their
disadvantage is that many interesting symmetries, like rotations among
discrete variables, are lost in this approach.

    A complementary approach was first developed for linear difference
equations \cite{11,7}, again given on fixed uniform lattices. It was formulated
in terms of linear difference operators, commuting with the linear
operator defining the original difference equation. This approach provides
a large number of symmetries and the symmetry algebras of the discrete
equations and their continuous limits are actually isomorphic.  The
symmetries of the difference equations are not point ones:  they act at
many points of the lattice. They do however provide flows that commute
with the flow determined by the original equation and can thus be used to
obtain solutions.

   This aspect of commuting flows has been adapted to nonlinear difference
and differential-difference equations \cite{8,lr,9, hlw, hlw1}. The equations are
defined on a fixed and uniform lattice. Generalized symmetries are
considered together with point ones and some of the generalized symmetries
reduce to point ones in the continuous limit. The methods for finding
these generalized symmetries rely on either linearizability, as in the
case of the discrete Burgers equation \cite{8}, or on integrability (the
existence of a Lax pair) as in the case of the Toda hierarchy \cite{lr,9,hlw}, or the discrete nonlinear Schr\"odinger equation \cite{hlw1,hl}.

This symmetry approach is powerful whenever it is applicable. Together with point and generalized symmetries it provides B\"acklund transformations as a composition of infinitely many higher symmetry transformations. This aspect has been explored in detail for the Toda lattice \cite{hlw}. We emphasize that B\"acklund transformations for difference equations, just as for differential ones, are not obtained directly as Lie symmetries (not even as generalized ones).

   Each of the above methods has its own merits and will be further
developed in the future.

   In this article we take the same point of view as in our recent article
\cite{1}. We consider point symmetries only and use the formalism of vector
fields acting on all variables, dependent and independent ones. In \cite{1}
we considered only one discretely changing variable. The lattice was not
fixed. Instead it was given by a further difference equation. Point symmetries
act on the entire difference system: the equation and the lattice. The
lattice is not necessarily uniform and we explored the effect of choosing
different types of lattices. The idea of using transforming lattices is
due to Dorodnitsyn and coworkers \cite{13,14,15,d,16}. We differ from them in one
crucial aspect. They start from a given symmetry group and construct
invariant difference schemes for a given group. We, on the other hand,
start from a given difference scheme and find its Lie point symmetry
group.  Previously this was done for the case of one independent variable.
In this article we generalize to the multidimensional case. The
generalization is by no means trivial. The lattice is given by $N^2$
equations, where $N$ is the number of independent variables, all of them
varying discretely. Transformations of continuously varying independent
variables, if present, are also taken into account.

  We stress that the approach of this article complements those of
previous ones. The results of \cite{4} and \cite{5} are obtained if we chose a
special form of the lattice (e.g. $x_{m+1} - x_m = h$ in the case of one
independent variable, where $h$ is a fixed, nontransforming constant). We purpusely avoid any use of integrability. Like Lie theory for differential equations, this approach is applicable to arbitrary differential systems, integrable or not.

 A general formalism for determining the symmetry
algebra is presented in Section~2. It generalizes the algorithm presented
earlier \cite{1} for ordinary difference equations to the case of several
independent variables. In Section~3 we apply the algorithm to a discrete linear
heat equation
which we consider on several different lattices, each providing its own
symmetries. Section~4 is devoted to difference equations on lattices that are
invariant under Lorentz transformations. In Section~5 we discuss two different
discrete Burgers equations, one linearizable, the other not. The lattices are
the
same in both cases, the symmetry algebras turn out to be different. Section~6
treats symmetries of differential-difference equations, i.e. equations involving
both discrete and continuous variables. Some conclusions are drawn in the final
Section~7.

\section{General symmetry formalism}

\subsection{The difference scheme}

For clarity and brevity, let us consider one scalar equation for a continuous
function of two (continuous) variables: $u=u(x,t)$. A lattice will be a set of
points $P_i$, lying in the plane $\mathbb{R}^2$ and stretching in all directions
with no boundaries. The points $P_i$ in $\mathbb{R}^2$ will be labeled by two
discrete labels $P_{m,n}$. The Cartesian coordinates of the point $P_{m,n}$
will be $(\xmn,\tmn)$ with $-\infty < m < \infty\ ,\ -\infty < n < \infty$
(we are of course not obliged to use Cartesian coordinates). The value of the
dependent variable in the point $P_{m,n}$ will be denoted $\umn=u(\xmn,\tmn)$.

A difference scheme will be a set of equations relating the values of
$\{x,t,u\}$ in a finite number of points. We start with one `reference point'
$P_{m,n}$ and define a finite number of points $P_{m+i,n+j}$ in the
neighborhood of $P_{m,n}$. They must lie on two different curves,
intersecting in $P_{m,n}$. Thus, the difference scheme will have the form
\be
\ba{c}
E_a\Big(\left\{ x_{m+i,n+j},t_{m+i,n+j},u_{m+i,n+j} \right\} \Big)=0
\ \ \ \ \ \ 1\le a \le 5
\\*[2ex]
-i_1 \le i \le i_2\ \ \ \ -j_1 \le j \le j_2\ \ \ \
i_1,i_2,j_1,j_2 \in \mathbb{Z}^{\ge 0}.
\label{eq:2.1}
\ea
\ee

The situation is illustrated on Figure~1. It corresponds to a lattice determined
by 6 points. Our convention is that $x$ increases as $m$ grows, $t$ increases as
$n$ grows (i.e. $\xmu-\xmn \eq h_1>0\ ,\ \tnu-\tmn \eq h_2>0$). The scheme on
Figure~1 could be used e.g. to approximate a differential equation of third
order in $x$, second in $t$.

\bigskip

\begin{center}
\begin{picture}(160,160)
\put(0,0){\vector(1,0){160}} \put(0,0){\vector(0,1){160}}

\put(80,-15){\makebox(0,0)[c]{Figure 1: Points on a lattice}}

\put(150,-5){\makebox(0,0){$x$}} \put(-5,150){\makebox(0,0){$t$}} \qbezier(15,15)(32,52)(55,85)
\qbezier(55,85)(90,132)(145,150) \qbezier(15,150)(29,148)(55,85) \qbezier(55,85)(80,38)(145,30)
\put(55,85){\makebox(0,0){$\bullet$}} \put(60,85){\makebox(0,0)[l]{$P_{m,n}$}}

\put(25,35){\makebox(0,0){$\bullet$}} \put(28,35){\makebox(0,0)[l]{$P_{m-1,n}$}}

\put(82,114){\makebox(0,0){$\bullet$}} \put(89,114){\makebox(0,0)[l]{$P_{m+1,n}$}}

\put(120,140){\makebox(0,0){$\bullet$}} \put(125,135){\makebox(0,0)[l]{$P_{m+2,n}$}}

\put(28,140){\makebox(0,0){$\bullet$}} \put(33,140){\makebox(0,0)[l]{$P_{m,n+1}$}}

\put(105,40){\makebox(0,0){$\bullet$}} \put(105,45){\makebox(0,0)[l]{$P_{m,n-1}$}}

\end{picture}
\setcounter{figure}{1}
\end{center}

\bigskip

\bigskip

Of the above five equations in (\ref{eq:2.1}), four determine the lattice, one
the difference equation. If a continuous limit exists, it is a partial
differential equation in two variables. The four equations determining the
lattice will reduce to identities (like $0=0$).

The system (\ref{eq:2.1}) must satisfy certain independence criteria. Starting
from the reference point $P_{m,n}$ and a given number of neighboring points, it
must be possible to calculate the values of $\{x,t,u\}$ in all points. This
requires a minimum of five equations: to be  able to calculate the $(x,t)$ in
two directions and $u$ in all points. For instance, to be able to move upward
and to the right along the curves passing through $P_{m,n}$ (with either $m$, or
$n$ fixed) we impose a condition on the Jacobian
\be
\left|J\right|=\left| \frac{\pa(E_1,E_2,E_3,E_4,E_5)}
{\pa(x_{m+i_2,n},t_{m+i_2,n},x_{m,n+j_2},t_{m,n+j_2},u_{m+i_2,n+j_2})}\right|
\ne 0.
\label{eq:2.2}
\ee

As an example of difference scheme, let us consider the simplest and most
standard lattice, namely a uniformly spaced orthogonal lattice and a difference
equation approximating the linear heat equation on this lattice. Equations
(\ref{eq:2.1}) in this case are:
\be
\xmu-\xmn=h_1\ \ \ \ \ \ \tmu-\tmn=0
\label{eq:2.3a}
\ee
\be
\xnu-\xmn=0\ \ \ \ \ \ \tnu-\tmn=h_2
\label{eq:2.3b}
\ee
\be
\frac{\unu-\umn}{h_2}=\frac{\umu-2\umn+\umd}{(h_1)^2}
\label{eq:2.4}
\ee
where $h_1$ and $h_2$ are constants.

The example is simple and the lattice and the lattice equations can be solved
explicitly to give
\be
\xmn=h_1 m + x_0\ \ \ \ \tmn=h_2 n + t_0.
\label{eq:2.5}
\ee

The usual choice is $x_0=t_0=0\ ,\ h_1=h_2=1$ and then $x$ is simply identified
with $m$, $t$ with $n$. We need the more complicated two index notation to
describe arbitrary lattices and to formulate the symmetry algorithm (see below).

The example suffices to bring out several points:

\begin{enumerate}

\item Four equations are needed to describe the lattice.

\item Four points are needed for equations of second order in $x$, first in $t$.
Only three figure in the lattice equation, namely $P_{m+1,n}, P_{m,n}$ and
$P_{m,n+1}$. To get the fourth point, $P_{m-1,n}$, we shift $m$ down by one unit
in equations (\ref{eq:2.3a}-\ref{eq:2.4}).

\item The independence condition (\ref{eq:2.2}) is needed to be able to solve
for $\xmu$, $\tmu,\xnu,\tnu$ and $\unu$.

\end{enumerate}

\subsection{Symmetries of the difference scheme}

We are interested in point transformations of the type
\be
\ti{x}=F_{\la}(x,t,u)\ \ \ \ \ti{t}=G_{\la}(x,t,u)\ \ \ \ \ti{u}=H_{\la}(x,t,u)
\label{eq:2.6}
\ee
where $\la$ is a group parameter, such that when $(x,t,u)$ satisfy the system
(\ref{eq:2.1}) then $(\ti x,\ti t, \ti u)$ satisfy the same system. The
transformation acts on the entire space $(x,t,u)$, at least locally, i.e. in
some neighborhood of the reference point $P_{m,n}$, including all points
$P_{m+i,n+j}$ figuring in equation (\ref{eq:2.1}). That means that the same
functions $F,G$ and $H$ determine the transformation of all points. The
transformations (\ref{eq:2.6}) are generated by the vector field
\be
\hat X=\xi(x,t,u)\pa_x + \tau(x,t,u)\pa_t + \phi(x,t,u)\pa_u.
\label{eq:2.7}
\ee

We wish to find the symmetry algebra of the system (\ref{eq:2.1}), that is the
Lie algebra of the local symmetry group of local point transformations. To do
this we must prolong the action of the vector field $\hat X$ from the reference
point $(\xmn,\tmn,\umn)$ to all points figuring in the system (\ref{eq:2.1}).
Since the transformations are given by the same functions $F,G$ and $H$ at all
points, the prolongation of the vector field (\ref{eq:2.7}) is obtained simply
by evaluating the functions $\xi, \tau$ and $\phi$ at the corresponding points.

In order words, we can write
\be
\ba{l}
pr\, \hat X= \sum_{m,n} \Big[ \xi(\xmn,\tmn,\umn)\pa_{\xmn} +
\tau(\xmn,\tmn,\umn)\pa_{\tmn}
\\*[2ex]
 \ \ \ \ \ \ \ \ \ \ \ \ \ \ \ \ \ \
+\phi(\xmn,\tmn,\umn)\pa_{\umn}\Big],
\ea
\label{eq:2.8}
\ee
where the summation is over all points figuring in the system (\ref{eq:2.1}).
The invariance requirement is formulated in terms of the prolonged vector field
as

\be
\mpr \hat X\, E_a\left|_{E_{b}=0}\ \ \ \ 1 \le a,b \le 5. \right.
\label{eq:2.9}
\ee

Just as in the case of ordinary difference equations, we can turn equation
(\ref{eq:2.9}) into an algorithm for determining the symmetries, i.e. the
coefficients in vector field (\ref{eq:2.7}).

The procedure is as follows:
\begin{enumerate}

\item Use the original equations (\ref{eq:2.1}) and the Jacobian condition
(\ref{eq:2.2}) to express five independent quantities in terms of the other
ones, e.g.
\be
\ba{c}
v_1=x_{m+i_2,n}\ \ \ \ v_2=t_{m+i_2,n}\ \ \ \ v_3=x_{m,n+j_2}
\\*[2ex]
v_4=t_{m,n+j_2}\ \ \ \ v_5=u_{m+i_2,n+j_2}
\ea
\label{eq:2.10}
\ee
as
\be
\ba{c}
v_a=v_a(x_{m+i,n+j}, t_{m+i,n+j}, u_{m+i,n+j})
\\*[2ex]
-i_1 \le i \le i_2-1\ \ \ \ -j_1 \le j \le j_2-1.
\ea
\label{eq:2.11}
\ee

\item Write the five equations (\ref{eq:2.9}) explicitly and replace the
quantities $v_a$ using equation (\ref{eq:2.11}). We obtain five functional
equations for the functions $\xi, \tau$ and $\phi$, evaluated at different point
of the lattice. Once the functions $v_a$ are substituted into these equations,
each value of $x_{i,k}, t_{i,k}$ and $u_{i,k}$ is independent. Moreover, it can
only figure via the corresponding $\xi_{i,k}, \tau_{i,k}$ and $\phi_{i,k}$ (with
the same values of $i$ and $k$), via the functions $v_a$, or explicitly via the
functions $E_a$.

\item Assume that the dependence of $\xi, \tau$ and $\phi$ on their variables is
analytic. Convert the obtained functional equations into a system of
differential equations by differentiating with respect to the variables
$x_{i,k}, t_{i,k}$ and $u_{i,k}$. This provides an overdetermined system of
linear partial differential equations which we must solve.

\item The solutions of the differential equations must be substituted back into
the functional ones and these in turn must be solved.
\end{enumerate}

The above algorithm provides us with the function $\xi(x,t,u)$, $\tau(x,t,u)$
and $\phi(x,t,u)$ figuring in equation (\ref{eq:2.7}). The finite
transformations of the (local) Lie symmetry group are obtained in the usual
manner, by integrating the vector field (\ref{eq:2.7}):
\be
\ba{ccc}
\frac{d\ti{x}}{d\la}=\xi(\ti x, \ti t, \ti u) &
\frac{d\ti{t}}{d\la}=\tau(\ti x, \ti t, \ti u) &
\frac{d\ti{u}}{d\la}=\phi(\ti x, \ti t, \ti u)
\\*[2ex]
\ti x\left|_{\lambda=0}\right.=x & \ti t\left|_{\lambda=0}\right.=t &
\ti u\left|_{\lambda=0}\right.=u.
\ea
\ee

\section{Discrete heat equation}

The heat equation in one-dimension
\be
u_t=u_{xx}
\label{eq:3.1}
\ee
is invariant under a six-dimensional Lie group, corresponding to translations in
$x$ and $t$, dilations, Galilei transformations, multiplication of u by a
constant and expansions. It is also invariant under an infinite dimensional
pseudo-group, corresponding to the linear superposition principle.

Symmetries of the discrete heat equation have been studied, using different
methods and imposing different restrictions on the symmetries \cite{7,11,13,d}.

Here we will use the discrete heat equation to illustrate the methods of
Section~2 and to show the influence of the choice of the lattice.

\subsection{Fixed rectangular lattice}
The discrete heat equation and a fixed lattice were given in equation
(\ref{eq:2.4}) and
(\ref{eq:2.3a}), (\ref{eq:2.3b}), respectively. Applying the operator
(\ref{eq:2.8}) to the lattice, we obtain
\bea
\xi(\xmu,\tmu,\umu)=\xi(\xmn,\tmn,\umn)
\label{eq:3.2a}
\\*[2ex]
\xi(\xnu,\tnu,\unu)=\xi(\xmn,\tmn,\umn).
\label{eq:3.2b}
\eea

The values $\umu, \unu, \umn$ are not related by equation (\ref{eq:2.4})
(since it also contains $\umd$). Hence if we differentiate equations
(\ref{eq:3.2a}), (\ref{eq:3.2b}), e.g. with respect to $\umn$, we find that
$\xi$ is independent of $u$. We have $\tmu=\tmn$ so equation (\ref{eq:3.2a})
implies that $\xi$ does not depend on $x$. Similarly, equation (\ref{eq:3.2b})
 implies that $\xi$ does not depend on $t$. Hence $\xi$ is constant.
Similarly, we obtain that $\tau(x,t,u)$ is also constant. Applying the
prolongation $pr \hat X$ to equation (\ref{eq:2.4}) we obtain the
functional equation
\be
\phi_{m,n+1}-\phi_{m,n}=\frac{h_2}{(h_1)^2}(\phi_{m+1,n}-2\phi_{m,n}+\phi_{m-1,n
})
\label{eq:3.3}
\ee
with e.g. $\phi_{m,n}\eq \phi(\xmn,\tmn,\umn)$.

In $\phi_{m,n+1}$ we replace $\unu$, using equation (\ref{eq:2.4}). We then
differentiate with respect to $\umu$ and again with respect to $\umd$. We obtain
\be
\phi_{m,n}=A(\xmn,\tmn)\umn + B(\xmn,\tmn).
\label{eq:3.4}
\ee
Substituting (\ref{eq:3.4}) into equation (\ref{eq:3.3}), using (\ref{eq:2.4})
again and setting the coefficient of $\umu, \umd, \umn$ and $1$ equal to zero
separately we find that $A$ must be constant and $B$ must be a solution of
equation (\ref{eq:2.4}). Thus, the symmetry algebra of the heat equation on the
lattice (\ref{eq:2.3a}), (\ref{eq:2.3b}) is given by
\be
\hat{P}_0=\pa_t\ \ \ \ \hat{P}_1=\pa_x\ \ \ \ \hat W=u\pa_u\ \ \ \
\hat S=S(x,t) \pa_u
\label{eq:3.5}
\ee
with $S$ a solution of the equation itself. Thus, the only symmetries are those
due to the fact that the equation is linear and autonomous.

\subsection{Lattices invariant under dilations}

There are at least two ways of making the discrete heat equation invariant under
dilations.

\subsection*{A) Five point lattice}
We replace the system of equations (\ref{eq:2.3a}), (\ref{eq:2.3b}) and
(\ref{eq:2.4}) by
\bea
\xmu-2\xmn+\xmd=0\ \ \ \ \ \ \ \ \ \xnu-\xmn=0
\label{eq:3.6a}
\\*[2ex]
\tmu-\tmn=0\ \ \ \ \ \ \ \ \ \tnu-2\tmn+\tnd=0
\label{eq:3.6b}
\\*[2ex]
\frac{\unu-\umn}{\tnu-\tmn}=\frac{\umu-2\umn+\umd}{(\xmu-\xmn)^2}.
\label{eq:3.7}
\eea

Applying $pr \hat X$ of equation (\ref{eq:2.7}) to (\ref{eq:3.6a}) and
substituting for $\xmu, \tmu, \tnu$ and $\xnu$ from the equations
(\ref{eq:3.6a}), (\ref{eq:3.6b}) we obtain
\be
\ba{l}
\xi(2\xmn-\xmd, \tmn, \umu)-2\xi(\xmn,\tmn, \umn)
\\*[2ex]
+\xi(\xmd,\tmd,\umd)=0
\label{eq:3.9a}
\ea
\ee
\be
\xi(\xmn, 2\tmn-\tnd, \unu)=\xi(\xmn,\tmn, \umn).
\label{eq:3.9b}
\ee

Since $\unu$ and $\umn$ are independent a differentiation of (\ref{eq:3.9b})
with respects to say $\umd$ (contained on the left hand side via $\unu$) implies
that $\xi$ does not depend on $u$. Differentiating (\ref{eq:3.9b}) with respect
to $\tnd$ we find that $\xi$ cannot depend on $t$ either. Putting $\xi=\xi(x)$
into equation (\ref{eq:3.9a}) and taking the second derivative with respect to
$\xmd$ and $\xmn$, we obtain that $\xi$ is linear in $x$. Similarly, invariance
of equation (\ref{eq:3.6b}) restricts the form of $\tau(x,t,u)$. Finally the
lattice (\ref{eq:3.6a}), (\ref{eq:3.6b}) is invariant under the transformation
generated by $\hat X$ with
\be
\xi=\al x + \b\ \ \ \ \tau=\ga t + \de.
\label{eq:3.10}
\ee

Now let us apply $pr \hat X$ to equation (\ref{eq:3.7}). We obtain
\be
\frac{\phi_{m,n+1}-\phi_{m,n}}{\tnu-\tmn}=\frac{\phi_{m+1,n}-2\phi_{m,n}+\phi_{m
-1,n}}{(\xmu-\xmn)^2}-(2\al-\ga)\frac{\umu-2\umn+\umd}{(\xmu-\xmn)^2}.
\label{eq:3.11}
\ee

Taking the second derivative $\pa_{\umu}\pa_{\umd}$ of equation (\ref{eq:3.11})
after using the equation (\ref{eq:3.7}) to eliminate $\unu$, we find
$\phi_{m,n}=A_{m,n}(x,t) \umn +
B_{m,n}(x,t)$. Substituting back into equation (\ref{eq:3.11}) we obtain
$A_{m,n}=A=const.$, and see that $B_{m,n}(x,t)$ must  satisfy the original
difference system. Moreover, we obtain the restriction $\ga=2\al$.

Finally, on the lattice (\ref{eq:3.6a}), (\ref{eq:3.6b}) the heat equation
(\ref{eq:3.7}) has a symmetry algebra generated by the operators (\ref{eq:3.5})
and the additional dilation operator
\be
\hat D=x\, \pa_x+2t\, \pa_t.
\label{eq:3.12}
\ee

We mention that the lattice equations (\ref{eq:3.6a}), (\ref{eq:3.6b}) can be
solved to give $x=am+b\ ,\ t=cn+d$. At first glance this seems to coincide with
the lattice (\ref{eq:2.5}). The difference is that in equation (\ref{eq:2.5})
$h_1$ and $h_2$ are fixed constants. Here $a,b,c$ and $d$ are integration
constants that can be chosen arbitrarily. In particular, they can be dilated.
Hence the additional dilational symmetry.

\subsection*{B) A four point lattice}

We only need four points to write the discrete heat equation, so it makes sense
to write a four point lattice. Let us define the lattice by the equations
\bea
\xmu-2\xmn+\xmd=0\ \ \ \ \ \ \ \ \ \xnu-\xmn=0
\label{eq:3.13a}
\\*[2ex]
\tmu-\tmn=0\ \ \ \ \ \ \ \ \ \tnu-\tmn-c(x_{m+1,n}-\xmn)^2=0.
\label{eq:3.13b}
\eea
On this lattice the discrete heat equation (\ref{eq:3.7}) simplifies to
\be
\unu-\umn=c(\umu-2\umn+\umd).
\label{eq:3.14}
\ee
Applying the same method as above, we find that invariance of the lattice
implies $\xi=Ax+B$, $\tau=2At+C$. Invariance of equation (\ref{eq:3.14}) then
implies $\phi=Du+S(x,t)$ where $A,B,C$ and $D$ are constants and $S(x,t)$ solves
the discrete heat equation. Thus, the discrete heat equation on the four point
lattice (\ref{eq:3.13a}), (\ref{eq:3.13b}) is invariant under the same group as
on the five point lattice (\ref{eq:3.6a}), (\ref{eq:3.6b}).

\subsection{Exponential lattice}

Let us now consider a lattice that is neither equally spaced, nor orthogonal,
given by the equations
\bea
\xmu-2\xmn+\xnd=0 & \xnu=(1+c)\, \xmn
\label{eq:3.15a} \\*[2ex]
\tnu-\tmn=h & \tmu-\tmn=0
\label{eq:3.15b}
\eea
with $c\ne 0,-1$. These equations can be solved and explicitly the lattice is
\be
t=hn+t_0\ \ \ \ x=(1+c)^n\, (\al m+\beta)
\label{eq:3.16}
\ee
where $t_0, \al$ and $\beta$ are integration constants. Thus while $t$ grows
by constant increments, $x$ grows with increments which vary exponentially with time (see Figure~2).
Numerically this
type of lattice may be useful if we can solve the equation asymptotically for
large values of $t$ and are interested in the small $t$ behavior.

\bigskip

\begin{figure}
\begin{center}
\includegraphics*[scale=0.3]{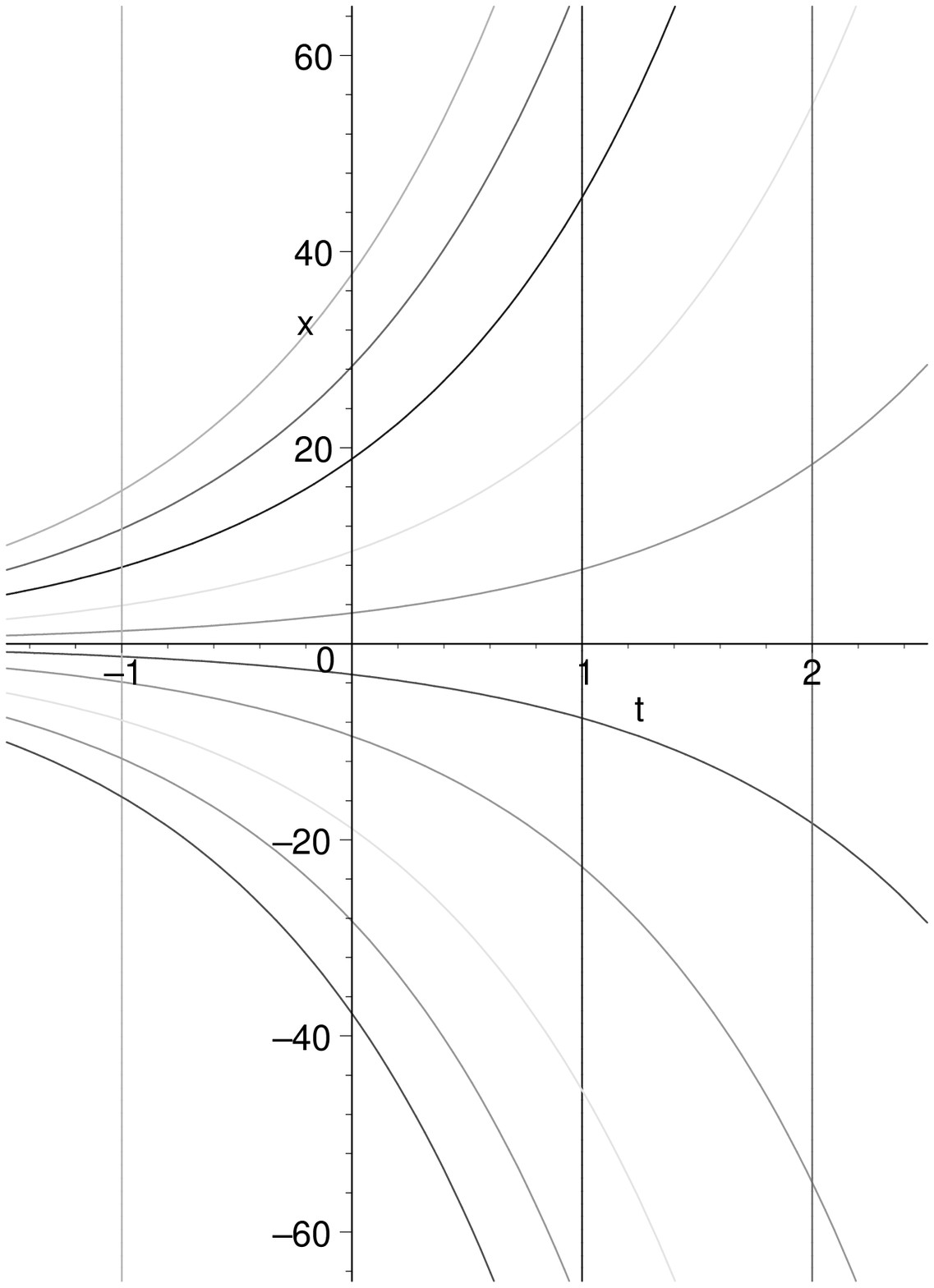}
\caption{Variables $(x,t)$ as functions of $m$ and $n$ for the lattice equations (\ref{eq:3.15a}),
(\ref{eq:3.15b}). The parameters and the integration constants are, respectively, $c=\sqrt{2}, h=1$ and
$\al=\pi, \beta=0, t_0=0$.} \label{fig:2}
\end{center}
\end{figure}

\bigskip
\bigskip

The heat equation on lattice (\ref{eq:3.15a}), (\ref{eq:3.15a}) can be written as \be
\frac{\unu-\umn}{h}=\frac{\umu-2\umn+\umd}{(\xmu-\xmn)^2}. \label{eq:3.17} \ee Applying the symmetry algorithm
to the lattice equations (\ref{eq:3.15a}), (\ref{eq:3.15b}) we find that the symmetry algebra is restricted to
\be \hat X=\left[a\, x+b\, (1+c)^{t/h}\right]\, \pa_x+\tau_0\, \pa_t+\phi(x,t,u)\, \pa_u, \label{eq:3.18} \ee
where $a, b$ and $\tau_0$ are arbitrary constants (whereas $c$ and $h$ are constants determining the lattice).
Invariance of the equation (\ref{eq:3.17}) implies $a=0$ in (\ref{eq:3.18}) and restricts $\phi(x,t,u)$ to
reflect linearity of the equation and nothing more. The resulting symmetry algebra has a basis consisting of \be
\hat{P}_1=(1+c)^{t/h}\, \pa_x\ \ \ \ \hat{P}_0=\pa_t\ \ \ \ \hat W=u\, \pa_u\ \ \ \ \hat S=S(x,t)\, \pa_u
\label{eq:3.19} \ee where $S(x,t)$ satisfies the heat equation. We see that the system is no longer invariant
under space translations, or rather, that these `translations' become time dependent and thus simulate a
transformation to a moving frame.

\subsection{Galilei invariant lattice}

Let us now consider the following difference scheme
\be
\frac{\unu-\umn}{\tau_2}=\tau_2^2\frac{\umu-2\umn+\umd}{\zeta^2}
\label{eq:3.20}
\ee
\be
\tmu-\tmn=\tau_1\ \ \ \ \ \ \tnu-\tmn=\tau_2
\label{eq:3.21a}
\ee
\be
\xmu-2\xmn+\xmd=0
\label{eq:3.21b}
\ee
\be
(\xmu-\xmn)\tau_2-(\xnu-\xmn)\tau_1=\zeta
\label{eq:3.21c}
\ee
where $\tau_1, \tau_2$ and $\zeta$ are fixed constants.

The lattice equations can be solved and we obtain
\be
\tmn=\tau_1\, m+\tau_2\, n+t_0\ \ \ \ \ \
\xmn=\si\tau_1\, m+ \left(\frac{\si \tau_1 \tau_2-\zeta}{\tau_1}\right)n+x_0
\label{eq:3.22}
\ee
where $\si, t_0$ and $x_0$ are integration constants. The corresponding lattice
is equally spaced and in general, nonorthogonal (see Figure~3). Indeed, the
coordinate curves, corresponding to $m=const$ and $n=const$, respectively, are
\be
\ba{rcl}
x-x_0&=&\si\, (t-t_0)-\frac{\zeta}{\tau_1}\, n
\\*[2ex]
x-x_0&=&\frac{\si\tau_1\tau_2-\zeta}{\tau_1\tau_2}\,
(t-t_0)+\frac{\zeta}{\tau_2}\,
m.
\label{eq:3.23}
\ea
\ee
These are two families of straight lines, orthogonal only in the special case
$(\si^2+1)\tau_1\tau_2=\si \zeta$. If we choose
\be
\si \tau_1 \tau_2-\zeta=0
\label{eq:3.24}
\ee
then the second family of coordinate lines in equation (\ref{eq:3.23}) is
parallel to the $x$~axis.

\begin{figure}
\begin{center}
\includegraphics*[scale=0.3]{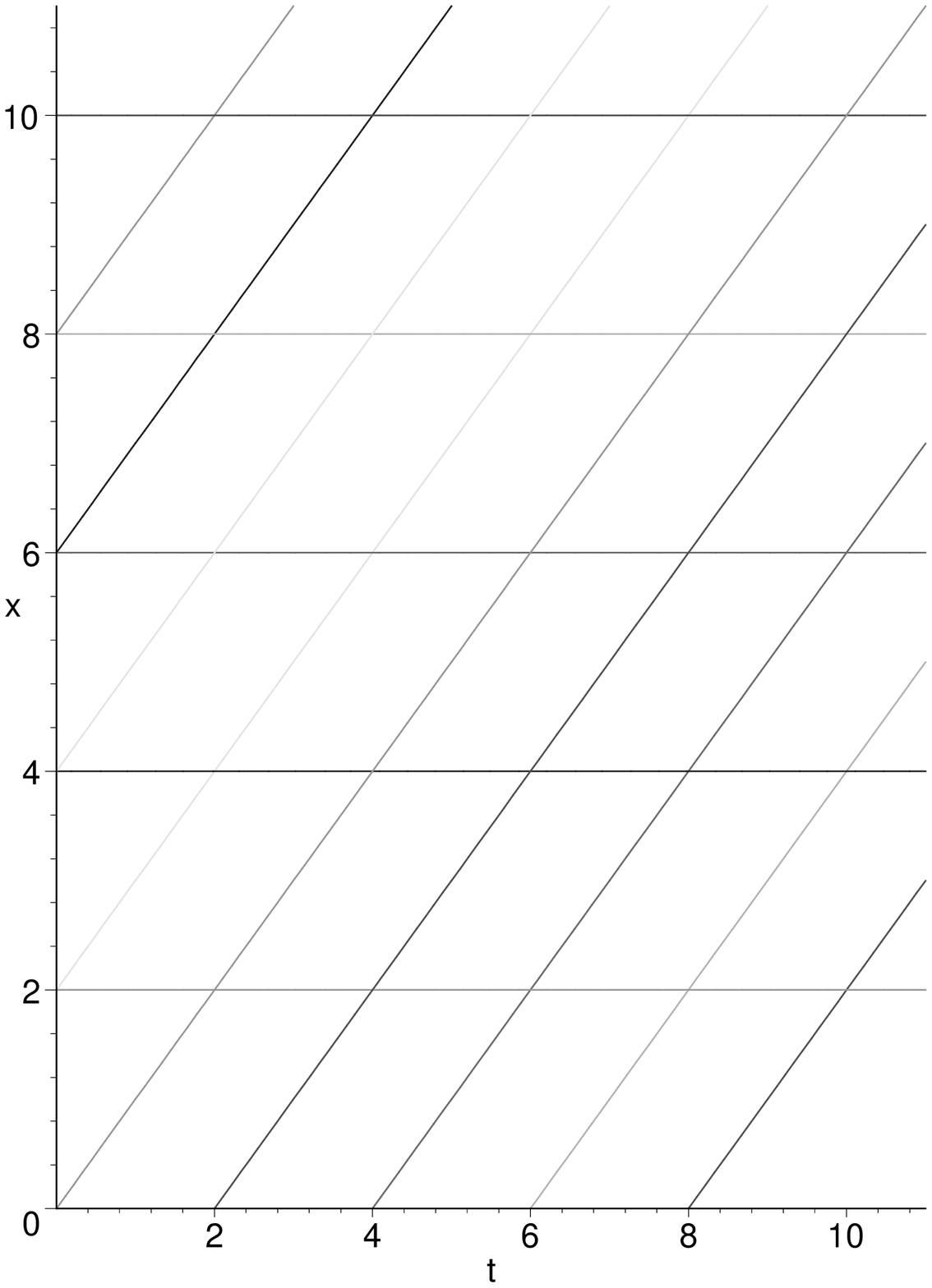}
\caption{Variables $(x,t)$ as functions of $m$ and $n$ for the lattice equations (\ref{eq:3.21a}),
(\ref{eq:3.21b}), (\ref{eq:3.21c}). The parameters and the integration constants are, respectively, $\tau_1=1,
\tau_2=2, \ze=2$ and $\si=1, x_0=0, t_0=0$.}
\end{center}
\end{figure}

Invariance of equation (\ref{eq:3.21a}) implies that in the vector field we have
$\tau(x,t,u)=\al=const$. From the invariance of equation (\ref{eq:3.21b}) we
obtain $\xi=A(t)\, x+B(t)$ with
\be
A(\tmu)=A(\tmn)\ \ \ \ \ \ B(\tmu)-2B(\tmn)+B(\tmd)=0.
\label{eq:3.25}
\ee
Finally, invariance of equation (\ref{eq:3.21c}) implies $A(t)=0$ and
$B(t)=\beta\, t+\ga$ where $\beta$ and $\ga$ are constants. Now let us apply the
prolonged vector field to
equation (\ref{eq:3.20}). We obtain $\phi=R\, u+S(x,t)$ where $S(x,t)$ satisfies
the system (\ref{eq:3.20}),...,(\ref{eq:3.21c}). The symmetry algebra is given
by

\be
\hat{P}_0=\pa_t\ \ \ \ \hat{P}_1=\pa_x\ \ \ \ \hat B=t\, \pa_x\ \ \ \ \hat W=u\,
\pa_u\ \ \ \ \hat S=S(x,t)\, \pa_u.
\label{eq:3.26}
\ee
Thus, the system is Galilei invariant with Galilei transformation generated by
the operator $\hat B$.

Let us now consider the continuous limit of the system
(\ref{eq:3.20}),...,(\ref{eq:3.21c}). We use the solution (\ref{eq:3.22}) of the
lattice equations (\ref{eq:3.21a}), (\ref{eq:3.21b}), (\ref{eq:3.21c}) and for
simplicity restrict the constants by imposing equation (\ref{eq:3.24}). We have,
from equation (\ref{eq:3.22}), (\ref{eq:3.24})
\be
\ba{cc}
\tnu=\tmn+\tau_2 & \xnu=\xmn
\\*[2ex]
x_{m\pm 1,n}=\xmn\pm \si\tau_1 & t_{m\pm 1,n}=\tmn\pm \tau_1.
\label{eq:3.27}
\ea
\ee
The continuous limit is obtained by pushing $\tau_1 \ll 1,\ \tau_2\ll 1,\ \ze
\ll 1$ and expanding both sides of equation (\ref{eq:3.20}) into a Taylor
series, keeping only the lowest order terms. The LHS of equation (\ref{eq:3.20})
gives
\[
\ba{rcl}
\frac{\unu-\umn}{\tau_2} &=& \frac{u(\xmn,\tmn+\tau_2)-u(\xmn,\tmn)}{\tau_2}
\\*[2ex]
&=& u_t+ \mathcal{O}(\tau_2)
\ea
\]
and the RHS is given by
\[
\ba{l}
\left(\frac{\tau_2}{\ze}\right)^2\, (\umu-2\umn+\umd)
\\*[2ex]
= \left(\frac{\tau_2}{\ze}\right)^2\, \left[
u(\xmn+\si\tau_1,\tmn+\tau_1)-2u(\xmn,\tmn)+u(\xmn-\si\tau_1,\tmn-\tau_1)\right]
\\*[2ex]
=u_{xx}+\frac{2}{\si}u_{x,t}+\frac{1}{\si^2}u_{tt}+\mathcal{O}(\tau_1).
\ea
\]
The continuous limit of the system (\ref{eq:3.20}),..., (\ref{eq:3.21c}) is
\be
u_t=u_{xx}+\frac{2}{\si}u_{x,t}+\frac{1}{\si^2}u_{tt}\ \ \ \ \si\ne 0.
\label{eq:3.28}
\ee
The symmetry algebra of this equation, for any value of $\si$, is isomorphic to
that of the heat equation. In addition to the pseudo-group of the superposition
principle, we have
\be
\ba{c}
\hat{P}_0=\pa_t\ \ \ \ \hat D=x\, \pa_x+2t\, \pa_t-\frac{1}{2}u\, \pa_u-cx\,
\pa_t
\\*[2ex]
\hat K=tx\, \pa_x+t^2\, \pa_t-\frac{1}{2}(t+\frac{1}{2}x^2)u\, \pa_u-c(x^2\,
\pa_x+xt\, \pa_t-\frac{1}{2}xu\, \pa_u)
\\*[2ex]
\hat{P}_1=\pa_x+c\, \pa_t\ \ \ \ \hat W=u\, \pa_u
\\*[2ex]
\hat B=t\, \pa_x-\frac{1}{2}xu\, \pa_u-c(x\, \pa_x-2t\, \pa_t)-c^2x\, \pa_t\ \ \
\ \ \ c\eq 1/\si.
\ea
\label{eq:3.29}
\ee
The fact that the commutation relations do not depend on $c$ suggest that
equation (\ref{eq:3.28}) could be transformed into the heat equation. This is
indeed the case and it suffices to put
\be
\ba{c}
u(x,t)=\me^{\frac{c[(2+c^2)x+ct]}{4(1+c^2)^2}}w(\al,\beta)
\\*[2ex]
\al=x+ct\ \ \ \ \ \beta=(1+c^2)(t-cx)
\ea
\label{eq:3.30}
\ee
to obtain
\be
w_{\beta}=w_{\al\al}.
\label{eq:3.31}
\ee
Notice that while the difference equation (\ref{eq:3.20}) on the lattice
(\ref{eq:3.21a}), (\ref{eq:3.21b}), (\ref{eq:3.21c}) is Galilei invariant, this
invariance is realized in a different manner, than for the continuous limit
(\ref{eq:3.28}). To see this, compare the operator $\hat B$ of equation
(\ref{eq:3.26}) with that of equation (\ref{eq:3.29}).

\section{Lorentz invariant equations}

The partial differential equation
\be
u_{xy}=f(u)
\label{eq:4.1a}
\ee
is invariant under the inhomogeneous Lorentz group, with its Lie algebra
realized as
\be
\hat{X}_1=\pa_x\ \ \ \ \hat{X}_2=\pa_y\ \ \ \ \hat{L}=y\pa_x-x\pa_y
\label{eq:4.2a}
\ee
(for any function $f(u)$). In equation (\ref{eq:4.1a}) $x$ and $y$ are `light
cone' coordinates. In the continuous case we can return to the usual space-time
coordinates $z=x+y$, $t=x-y$, in which we have
\be
u_{zz}-u_{tt}=f(u)
\label{eq:4.1b}
\ee
instead of equation (\ref{eq:4.1a}) and the Lorentz group is generated by
\be
\hat{P}_0=\pa_t\ \ \ \ \hat{P}_1=\pa_z\ \ \ \ \hat{L}=t\pa_z+z\pa_t.
\label{eq:4.2b}
\ee

Let us now consider a discrete system, namely
\be
\frac{u_{m+1,n+1}-u_{m,n+1}-u_{m+1,n}+u_{m,n}}{(\xmu-\xmn)(y_{m,n+1}-y_{m,n})}=f
(\umn)
\label{eq:4.3}
\ee
\bea
\xmu-2\xmn+\xmd=0 && \xnu-\xmn=0
\label{eq:4.4a}
\\*[2ex]
\ynu-2\ymn+\ynd=0 && \ymu-\ymn=0.
\label{eq:4.4b}
\eea
Applying the operator $pr \hat{X}$ (with $t$ replaced by $y$) of equation
(\ref{eq:2.8}) to equations (\ref{eq:4.4a}), (\ref{eq:4.4b}) we obtain
\be
\xi=Ax+C\ \ \ \ \eta=By+D.
\label{eq:4.5}
\ee
Requesting the invariance of equation (\ref{eq:4.3}) we find that $\phi$ must be
linear
\be
\phi=\al(x,y)u+\b(x,y).
\label{eq:4.6}
\ee
The remaining determining equations yield $\al=\al_0=const.$ and
\be
(A+B)\frac{\pa f}{\pa \umn}+\left(\al_0 \umn + \b(x,y)\right)\frac{\pa^2 f}{\pa
\umn^2}=0.
\label{eq:4.7}
\ee

Thus, for any function $f=f(u)$ we obtain the symmetries (\ref{eq:4.2a}), just
as in the continuous case (they correspond to $B=-A$, $\al_0=\b=0$). As in the
continuous case, the symmetry algebra can be larger for special choices of the
function $f(u)$. Let us analyze these cases.

\subsection*{a) Nonlinear interaction}
We have $f''\ne 0$, hence $\b=\b_0=const$. The function must then satisfy
\be
(A+B-\al_0)f+(\al_0u+\b)f'=0.
\label{eq:4.8}
\ee
For $\al_0\ne 0$ we take
\be
f=u^p\ \ \ \ p\ne 0,1
\label{eq:4.9a}
\ee
(we have dropped some inessential constants). The system (\ref{eq:4.3}),
(\ref{eq:4.4a}), (\ref{eq:4.4b}) is, in this case,
invariant under a four-dimensional group generated by the algebra
(\ref{eq:4.2a}), complemented by dilation
\be
\hat D=x\pa_x+y\pa_y+\frac{2}{1-p}u\pa_u.
\label{eq:4.9b}
\ee

For $\al_0=0$, $\b \ne 0$ we have
\be
f=\me^u.
\label{eq:4.10}
\ee
The algebra is again four-dimensional with the additional dilation
\be
\hat D=x\pa_x+y\pa_y-2\pa_u.
\label{eq:4.11}
\ee

\subsection*{b) Linear interaction $f(u)=u$}
The only elements of the Lie algebra additional to (\ref{eq:4.2a}) are
\be
\hat D=u\pa_u\ \ \ \ \hat{S}(\b)=\b\pa_u
\label{eq:4.12}
\ee
where $\b$ satisfies the system(\ref{eq:4.3}), (\ref{eq:4.4a}), (\ref{eq:4.4b})
with $f(u)=u$. The presence of $\hat D$ and $\hat{S}(\b)$ is just a consequence
of linearity.

\subsection*{c) Constant interaction $f(u)=1$}
The additional elements of the Lie algebra are again a consequence of linearity,
namely
\be
\hat L=x\pa_x+y\pa_y+2u\pa_u\ \ \ \ \hat S=[S_1(x)+S_2(y)]\pa_u
\label{eq:4.13}
\ee
where $S_1(x)$ and $S_2(y)$ are arbitrary (because $S_1(x)+S_2(y)$ is the
general solution of equation (\ref{eq:4.3}) with $f(u)=0$ on the
lattice (\ref{eq:4.4a}), (\ref{eq:4.4b})).

To find a discretization of equation (\ref{eq:4.1b}), invariant under the group
corresponding to (\ref{eq:4.2b}) is more difficult and we will not go into that
here.

As stressed in the Introduction, the methods of this article can be applied to any difference system, but they provide only point symmetries.  We could treat the integrable discrete Liouville and Sine-Gordon equations of Faddeev \cite{f}, or Hirota \cite{hi}, but would not otain the generalized symmetries that are of interest. The correct formalism to use for these equations is that of Ref. \cite{9}.

\section{Discrete Burgers equation}

The continuous Burgers equation is written as
\be
u_t=u_{xx}+2uu_x,
\label{eq:5.1}
\ee
or in potential form as
\be
v_t=v_{xx}+v_{x}^{2}\ \ \ \ \ \ u\eq v_x.
\label{eq:5.2}
\ee

We shall determine the symmetry groups of two different discrete Burgers
equations, both on the same lattice. The lattice is one of those used above for
the heat equation, namely the four point lattice (\ref{eq:3.13a}),
(\ref{eq:3.13b}). Each of the four lattice equations involves at most three
points. Hence, for any difference equation on this lattice, involving all four
points, the symmetry algebra will be realized by vector fields of the form
(\ref{eq:2.7}) with
\be
\xi=Ax+B\ \ \ \ \ \ \tau=2At+D
\label{eq:5.3}
\ee
where $A, B$ and $D$ are constants (see section~3.2B).

\subsection{Nonintegrable discrete potential Burgers equation}

An absolutely straightforward discretization of equation (\ref{eq:5.2}) on the
lattice (\ref{eq:3.13a}), (\ref{eq:3.13b}) is
\be
\frac{\unu-\umn}{\tnu-\tmn}=\frac{\umu-2\umn+\umd}{(\xmu-\xmn)^2}+
\left(\frac{\umu-\umn}{\xmu-\xmn}\right)^2.
\label{eq:5.4}
\ee

Applying the usual symmetry algorithm, we find a four-dimensional symmetry
algebra
\be
\hat{P}_1=\pa_x\ \ \ \ \hat{P}_0=\pa_t\ \ \ \ \hat D=x\, \pa_x+2t\, \pa_t\ \ \ \
\hat W=\pa_u.
\label{eq:5.5}
\ee

\subsection{A linearizable discrete Burgers equation}

A different discrete Burgers equation was proposed recently \cite{8}. It is
linearizable by a discrete version of the Cole-Hopf transformation. Using the
notation of this article, we write the linearizable equation as
\be
\unu=\umn
+c\, \frac{(1+h_x\umn)[u_{m+2,n}-2\umu+\umn+h_x\umu(u_{m+2,n}-\umn)}
{1+ch_x[\umu-\umn+h_x\umn\umu]}
\label{eq:5.6}
\ee
\[
\ba{cc}
h_x\eq\xmu-\xmn & h_t\eq\tnu-\tmn=ch_{x}^2
\\*[2ex]
\tmu-\tmn=0 & \xnu-\xmn=0.
\ea
\]
In equation (\ref{eq:5.6}) $c$ is a constant, but $h_x$ is a variable, subject
to dilations. The determining equation is obtained in the usual manner. It
involves the function $\phi_{m,n}$ at all points figuring in equation
(\ref{eq:5.6}), and also the constant $A$ of equation (\ref{eq:5.3}). The
equation is too long to be included here, but is straightforward to obtain. The
variable that we choose to eliminate using equation (\ref{eq:5.6}) is $\unu$.
Differentiating twice with respect to $u_{m+2,n}$ we obtain
\be
\frac{\pa^2\phi_{m,n+1}}{\pa\unu^2}\, \frac{\pa\unu}{\pa u_{m+2,n}}=
\frac{\pa^2\phi_{m+2,n}}{\pa u_{m+2,n}^2}.
\label{eq:5.7}
\ee
We differentiate (\ref{eq:5.7}) with respect to $\umn$ and then, separately,
 with respect to $\umd$. We obtain two equations that are compatible for
$c(1+c)^2 h_x (1+h_x \umn)=0$. Otherwise they imply that $\phi$ is linear in
$u$: $\phi=\al(x,t)\, u+\beta(x,t)$. We have $c\ne 0$, $h_x \ne 0$, but the case
$c=-1$ must be considered separately. We first introduce the expression for
$\phi$ into the determining equation and obtain, after a lengthy computation
(using MAPLE): $\al=-A$, $\beta=0$. For $c=-1$ we proceed differently, but got
the same result. Finally, the Lie point symmetry algebra of the system
(\ref{eq:5.6}), (\ref{eq:3.13a}), (\ref{eq:3.13b}) has the basis
\be
\hat{P}_0=\pa_t\ \ \ \ \hat{P}_1=\pa_x\ \ \ \ \hat D=x\, \pa_x+2t\, \pa_t-u\,
\pa_u.
\label{eq:5.8}
\ee

This result should be compared with the symmetry algebra of equation
(\ref{eq:5.6}) on a fixed constant lattice, found earlier \cite{8}. The
symmetry algebra found there was five-dimensional. It was inherited from the
heat equation, via the discrete Cole-Hopf transformation. It was realized in a
`discrete evolutionary formalism' by flows, commuting with the flow given by the
Burgers equation. The symmetries found there were higher symmetries, and cannot
be realized in terms of the vector fields of the form considered in this
article.

\section{Symmetries of differential-difference equations}

\subsection{General comments}

Symmetries of differential-difference equations were discussed in our previous
article \cite{1}. Here we shall put them into the context of partial difference
equations and consider a further example. As in the case of multiple discrete
variables, we will consistently consider the action of vector fields at points
in the space of independent and dependent variables. To do this we introduce a
discrete independent variable $n$ (or several such variables) and a continuous
independent variable $\al$ (or a vector variable $\vec{\al}$). A point in the
space of independent variables will be $P_{n,\al}$, its coordinates $\{x_{n,
\al}, z_{n,\al}\}$ where both $x$ and $z$ can be vectors. The form of the
lattice is specified by some relations between $x_{n, \al}, z_{n,\al}$ and
$u_{n,\al}\eq u(x_{n, \al}, z_{n,\al})$.

We shall not present the general formalism here, but restrict to the case of one
discretely varying variable $z\eq z_n$, $-\infty< n < \infty$ and either one
continuous (time) variable $(t)$, or two continuous variables $(x,y)$.

For instance, a uniform lattice that is time independent can be given by the
relations
\bea
z_{n+1,\al}-2z_{n,\al}+z_{n-1,\al}&=&0
\label{eq:6.1a}\\*[2ex]
z_{n,\al}-z_{n,\al'}&=&0
\label{eq:6.1b} \\*[2ex]
t_{n+1,\al}-t_{n,\al}&=&0.
\label{eq:6.1c}
\eea
where $\al'$ is a different value of the continuous variable $\al$.

Conditions (\ref{eq:6.1b}),  (\ref{eq:6.1c}) are rather natural. They state that
time is the same at each point of the lattice and that the lattice does not
evolve in time. They are however not obligatory. Similarly, equation
(\ref{eq:6.1a}) is not obligatory. The solution of equations (\ref{eq:6.1a}),
..., (\ref{eq:6.1c}) is of course trivial, namely
\be
z_n=h\, n+z_0\ \ \ \ \ \ t=t(\al)
\label{eq:6.2}
\ee
and we can identify $t$ and $\al$ ($t=\al$, $h$ and $z_0$ are constants).

The prolongation of a vector field acting on a differential-difference scheme on
the lattice (\ref{eq:6.1a}),..., (\ref{eq:6.1c}) will have the form
\be
\ba{rl}
pr\, \hat{X}= & \sum_{n} \Big[ \tau(z_{n,\al},t_{n,\al},u_{n,\al})\,
\pa_{t_{n,\al}}+
\ze(z_{n,\al},t_{n,\al},u_{n,\al})\, \pa_{z_{n,\al}}
\\*[2ex]
& +\phi(z_{n,\al},t_{n,\al},u_{n,\al})\, \pa_{u_{n,\al}}\Big]+ \ldots
\ea
\label{eq:6.3}
\ee
where the dots signify terms acting on time derivatives of $u$.
Since $u_{n,\al}$, $u_{n,\al'}$ and $u_{n+1,\al}$ are all independent,
equations (\ref{eq:6.1b}) and (\ref{eq:6.1c}) imply
\be
\ze=\ze(z_n)\ \ \ \ \ \ \tau=\tau(t).
\label{eq:6.4}
\ee
On any lattice satisfying equation (\ref{eq:6.1b}), (\ref{eq:6.1c}) we can
simplify notation and write
\be
\hat X=\ze(z)\, \pa_z+\tau(t)\, \pa_t+\phi(z,t,u)\, \pa_u.
\label{eq:6.5}
\ee

Similarly for an equation with one discretely varying independent variable $z$
and two continuous ones $(x,y)$ one can impose
\be
z_{n+1,\al_1,\al_2}-2z_{n,\al_1,\al_2}+z_{n-1,\al_1,\al_2}=0
\label{eq:6.6a}
\ee
\be
\ba{rcl}
z_{n,\al_1 ',\al_2}-z_{n,\al_1,\al_2}&=&0
\\
z_{n,\al_1,\al_2 '}-z_{n,\al_1,\al_2}&=&0
\ea
\label{eq:6.6b}
\ee
\be
\ba{rcl}
x_{n+1,\al_1,\al_2}-x_{n,\al_1,\al_2}&=&0
\\
y_{n+1,\al_1,\al_2}-y_{n,\al_1,\al_2}&=&0.
\ea
\label{eq:6.6c}
\ee
Invariance of the conditions (\ref{eq:6.6b}) and (\ref{eq:6.6c}) then implies
that the vector fields realizing the symmetry algebra have the form
\be
\hat X=\ze(z)\, \pa_z+\xi(x,y)\, \pa_x+\eta(x,y)\, \pa_y+\phi(z,x,y,u)\, \pa_u.
\label{eq:6.7}
\ee

We can again simplify notation identifying $x=\al_1$, $y=\al_2$ and solving
(\ref{eq:6.6a}) to give $z_n=h\, n+z_0$ ($h$ and $z_0$ constant).

\subsection{Examples}

We shall consider here just one example that brings out the role of the lattice
equations very clearly. The example is Toda field theory, or the two-dimensional
Toda lattice \cite{17,18,19}. It is given by the equation
\be
u_{n,xy}=\me^{u_{n-1}-u_n}-\me^{u_{n}-u_{n+1}}
\label{eq:6.8}
\ee
with $u_n\eq u(z_n,x,y)$.

On the lattice (\ref{eq:6.6a}),..., (\ref{eq:6.6c}) we start with equation
(\ref{eq:6.7}) and have
\be
pr\, \hat X=\xi(x,y)\, \pa_x+\eta(x,y)\, \pa_y+
\sum_{k=-1}^{1}\ze_{n+k}(z)\, \pa_{z_{n+k}}+
\sum_{k=-1}^{1}\phi_{n+k}\, \pa_{u_{n+k}}+
\phi_{n}^{xy}\, \pa_{u_{n,xy}}
\label{eq:6.9}
\ee
where $\phi_{n}^{xy}$ is calculated in the same way as for differential
equations \cite{2}.

Applying (\ref{eq:6.9}) to equations (\ref{eq:6.6a}) and (\ref{eq:6.8}) we find
\be
\xi=\xi(x)\ \ \ \ \eta=\eta(y)\ \ \ \ \ze_n=A\, z_n+B\ \ \ \
\phi_n=\beta_n(x,y,z_n)
\label{eq:6.10}
\ee
and we still have two equation to solve, namely
\bea
\beta_{n+1}-\beta_n+\xi_x+\eta_y&=&0
\label{eq:6.11} \\*[2ex]
\beta_{n,xy}&=&0.
\label{eq:6.12}
\eea
On the lattice (\ref{eq:6.6a}),..., (\ref{eq:6.6c}) $z_{n+1}$ and $z_n$ are
independent. Hence we can differentiate (\ref{eq:6.11}) with respect to
$z_{n+1}$ and find that $\beta_{n+1}$ is independent of $z_{n+1}$ and hence of
$n$. We thus find a symmetry algebra generated by
\be
\ba{c}
\hat{P}_1=\pa_x\ \ \ \ \hat{P}_2=\pa_y\ \ \ \ \hat L=x\, \pa_y-y\, \pa_x\ \ \ \
\hat S=\pa_z\ \ \ \ \hat D=z\, \pa_z
\\*[2ex]
\hat{U}(k)=k(x)\, \pa_u\ \ \ \ \hat{V}(h)=h(y)\, \pa_u
\ea
\label{eq:6.13}
\ee
where $k(x)$ and $h(y)$ are arbitrary smooth functions. Notice that $\hat S$
and $\hat D$ act only on the lattice and $\hat{U}(k)$ and $\hat{V}(h)$ generate
gauge transformations, acting only on the dependent variables.

If we change the lattice to a fixed, nontransforming one, i.e. replace
(\ref{eq:6.6a}) by
\be
z_{n+1,\al_1,\al_2}-z_{n,\al_1,\al_2}=h
\label{eq:6.14}
\ee
$h=const$, the situation changes dramatically.
We loose the dilation $\hat D$ of equation (\ref{eq:6.13}), however $z_{n+1}$
and $z_n$ are now related by equation (\ref{eq:6.14}). The solution of equation
(\ref{eq:6.11}), (\ref{eq:6.12}) in this case is
\be
\beta_n=\frac{z}{h}(\xi_x+\eta_y)+k(x)+h(y).
\label{eq:6.14b}
\ee
On this fixed lattice the Toda field equations are conformally invariant and the
invariance algebra is spanned by
\be
\ba{c}
\hat{X}(f)=f(x)\, \pa_x+\frac{z}{h}f'(x)\, \pa_u\ \ \ \
\hat{Y}(g)=g(y)\, \pa_y+\frac{z}{h}g'(y)\, \pa_u
\\*[2ex]
\hat{U}(k)=k(x)\, \pa_u\ \ \ \ \hat{V}(h)=h(y)\, \pa_u\ \ \ \ \hat{S}=\pa_z.
\ea
\label{eq:6.15}
\ee

We see that giving more freedom to the lattice (three points $z_{n+1}, z_n,
z_{n-1}$ instead of two) may lead to a reduction of the symmetry group, rather
than to an enhancement. For the Toda field theory the reduction is a drastic
one: the two arbitrary functions $f(x)$ and $g(y)$ reduce to $f=ax+b$,
$g=-ay+d$, respectively (and only the element $\hat D$ is added to the symmetry
algebra).

\section{Conclusions and future outlook}

The main conclusion is that we have presented an algorithm for determining the
Lie point symmetry group of a difference system, i.e. a difference equation and
the lattice it is defined on. The algorithm provides us with all Lie point
symmetries of the system. In Ref. \cite{1} we considered only one discretely varying
independent variable. In this article we concentrated on the case of two such
variables. The case of an arbitrary number of dependent and independent
variables is completely analogous though it obviously involves more cumbersome
notations and lengthier calculations. The problem of finding the symmetry group
is reduced to solving linear functional equations. In turn, these are converted
into an overdetermined system of linear partial difference equations, just as
in the case of differential equations. The fact that the determining equations
are linear, even if the the studied equations are nonlinear, is due to the
infinitesimal approach.

The symmetry algorithm can be computerized, just as it has been for differential
equations.

In previous articles (other than Ref. \cite{1}) we considered only one discretely
varying variable and a fixed (nontransforming) lattice \cite{4,5,10,6,7,8,lr,9,glw,19,lwm,hlw,ltw,hlw1,hl}. The
coefficients in the vector fields, realizing the symmetry algebra, depended on
variables evaluated at more than one point of the lattice, possibly infinitely
many ones. Thus, one obtained generalized symmetries together with point ones.
For integrable equations, including linear and linearizable ones, the symmetry
structure can be quite rich \cite{7,8,lr,9,hlw,hlw1,hl}. In the continuous limit some of the
generalized symmetries reduce to point ones \cite{9,hlw1,hl} and the structure of the
symmetry algebra changes.

A detailed comparison of various symmetry methods is postponed to a future
article. Applications of Lie point symmetries, as well as generalized
symmetries, to the solution of difference equations, will be given elsewhere.

\section*{Acknowledgments}

The work reported in this article was performed while S.T. and P.W. were
visiting the Dipartimento di Fisica of the Universit\`a di Roma Tre in the framework of the Universit\'e de
Montr\'eal -- Universit\`a Roma Tre interchange agreement. They thank this University and INFN for
hospitality and support. The research of P.W. is partly supported by grants from NSERC of Canada and FCAR of Qu\'ebec.

\end{document}